\documentclass[a4paper,fleqn,usenatbib]{mnras}
\usepackage[T1]{fontenc}
\usepackage{ae,aecompl}

\usepackage{amsmath}
\usepackage{graphicx}
\usepackage{lscape}

\usepackage{amssymb}
\usepackage{color}
\usepackage[flushleft]{threeparttable}

\newcommand {\bc}{\begin {center}}
\newcommand {\ec}{\end {center}}
\newcommand {\be}{\begin {equation}}
\newcommand {\ee}{\end {equation}}
\newcommand {\beq}{\begin {eqnarray}}
\newcommand {\eeq}{\end {eqnarray}}

\def\flux{erg s$^{-1}$ cm$^{-2}$}
\def\lum{erg s$^{-1}$}

\def\gx{GX~304$-$1}

\title[Spectral transition in \gx]
{Dramatic spectral transition of X-ray pulsar \gx ~ in low luminous state}

\author[S.~S.~Tsygankov et al.]
{Sergey~S.~Tsygankov,$^{1,2}$\thanks{E-mail: sergey.tsygankov@utu.fi} 
Alicia Rouco Escorial,$^{3}$
Valery F. Suleimanov,$^{4,5,2}$\newauthor 
Alexander A.~Mushtukov,$^{6,2,3}$
Victor Doroshenko,$^{4}$
Alexander A. Lutovinov,$^{2}$\newauthor 
Rudy Wijnands$^{3}$
and Juri~Poutanen$^{1,2}$
 \\
$^1$ Department of Physics and Astronomy,
  FI-20014 University of Turku, Finland \\
$^2$ Space Research Institute of the Russian Academy of Sciences, Profsoyuznaya Str. 84/32, Moscow 117997, Russia \\
$^3$ Anton Pannekoek Institute of Astronomy, University of Amsterdam, Science Park 904, 1098 XH Amsterdam, The Netherlands\\
$^4$ Institut f\"ur Astronomie und Astrophysik, Universit\"at T\"ubingen, Sand 1, D-72076 T\"ubingen, Germany \\
$^5$ Kazan (Volga region) Federal University, Kremlevskaya str.
 18, 420008 Kazan, Russia \\
$^6$ Leiden Observatory, Leiden University, NL-2300RA Leiden, the Netherlands}

\date{Accepted 2018 December 13. Received 2018 December 13; in original form 2018 October 22}

\pubyear{2018}

\begin{document}
\label{firstpage}
\pagerange{\pageref{firstpage}--\pageref{lastpage}}
\maketitle

\begin{abstract}
 We report on the discovery of a dramatic change in the energy spectrum of the X-ray pulsar \gx\ appearing at low luminosity. Particularly, we found that the cutoff power-law spectrum typical for accreting pulsars, including \gx\ at higher luminosities of $L_{\rm X}\sim 10^{36} - 10^{37}$ \lum, transformed at lower luminosity of $L_{\rm X}\sim 10^{34}$ \lum\ to a two-component spectrum peaking around 5 and 40~keV. We suggest that the observed transition corresponds to a change of the dominant mechanism responsible for the deceleration of the accretion flow. We argue that the accretion flow energy at low accretion rates is released in the  atmosphere of the neutron star, and the low-energy component in the source spectrum corresponds to the thermal emission of the optically thick, heated atmospheric layers. The most plausible explanations for the high-energy component are either the cyclotron emission reprocessed by the magnetic Compton scattering or the thermal radiation of deep atmospheric layers partly Comptonized in the overheated upper layers. Alternative scenarios are also discussed.
\end{abstract}


\section{Introduction}
\label{intro}
 Spectral energy distribution of astrophysical objects reflects physical processes responsible for the observed emission. Extreme conditions in the vicinity of highly magnetized neutron stars (NS), X-ray pulsars (XRPs), result in a very complex interplay between various physical and geometrical effects happening in the emission regions. As a consequence, no physically motivated spectral model able to describe
spectra of various XRPs at different luminosities emerged so far.
Moreover, the existing models focus on the description of X-ray spectrum at high luminosities \citep[see e.g.][]{2007ApJ...654..435B, 2016A&A...591A..29F} as no high-quality observations at low luminosities were available until recently.
From observational point of view, spectra of all bright XRPs are quite similar, and can be roughly described with a cutoff power-law type continuum.
We note, however, that there is at least one exception: a low-luminosity XRP X~Persei. 

The spectrum of this persistent Be X-ray binary deviates significantly from the standard picture and exhibits two distinct humps with maxima around 10 and 60~keV. Several interpretations have been suggested to describe the observed spectrum. \citet{1998ApJ...509..897D} proposed the continuum model consisting of two cutoff power-law components. The soft component was attributed to the partially thermalized emission from the NS atmosphere, whereas the hard component was suggested to be associated to the optically thin cyclotron emission in a strong magnetic field. 
Alternative interpretation was proposed by \citet{2012A&A...540L...1D}, who associated the two spectral components with thermal and bulk Comptonization in the vicinity of the NS. Finally, the source spectrum can be described with a single continuum component, for example, in the form of a power law modified by the high-energy cutoff, which is typical for bright XRPs. In this case, however, inclusion of a deep broad absorption feature at $\sim30$ keV is required \citep[cyclotron resonant scattering feature, CRSF; see e.g.][]{2001ApJ...552..738C,2012MNRAS.423.1978L}. 

The main difference between X~Persei and others XRPs observed in broad energy band is that it has a significantly lower luminosity of a few times $10^{34}$ \lum. 
Unfortunately, X~Persei is a persistent source with insignificant variations in the mass accretion rate,
so it was not possible  to observe this source also at higher luminosities, and verify whether its spectrum at much higher accretion rates is more similar to that of other XRPs.
On the other hand, observations of more distant sources at comparably low luminosity levels were not possible until recently, thus their behaviour at low accretion rates remained largely unexplored.

The situation has changed with the launch of {\it NuSTAR} observatory \citep{2013ApJ...770..103H}, which allows detailed investigations of transient systems with Be companions (Be/XRP) known to exhibit giant outbursts covering luminosity range of up to 5 orders of magnitude \citep[see e.g.][]{2011Ap&SS.332....1R}.

In this work we present results of the spectral analysis of the X-ray emission from one of the most suitable for such research transient Be/XRP \gx\ observed with \textit{NuSTAR} at a luminosity around $10^{34}$~\lum, the lowest ever observed for this source \citep{2018arXiv181101453R}.  The source exhibits regular Type I outbursts almost every periastron passage \citep{2017MNRAS.471.1553K}, covering broad range of mass accretion rates.  It  has a relatively long pulse period of $\sim272$~s \citep{1977ApJ...216L..15M} and orbital period of $\sim132.5$~d \citep{1983ApJ...273..709P}. The magnetic field of the NS in the system is known from the CRSF energy of $\sim54$ keV \citep{2011PASJ...63S.751Y}. Note that the CRSF energy also exhibits clear positive correlation with the source luminosity \citep{2012A&A...542L..28K}. The distance to the source is now known with high accuracy thanks to the {\it Gaia} results \citep[$d=2.01\pm0.15$ kpc,][]{2018arXiv180611397T}.

\section{Data analysis and results}
\label{sec:data}

The source was observed with \textit{NuSTAR} on 2018 June
3 (ObsID 90401326002; MJD 58272-58273). 
The raw data were processed following the standard data reduction
procedures described in the {\it NuSTAR} user guide, and using the standard {\it NuSTAR} Data Analysis Software ({\sc nustardas}) v1.6.0 provided under {\sc heasoft v6.24} with the CALDB version 20180814.  
The source and background spectra were extracted from the circular regions with radius of 60\arcsec\ using the {\sc nuproducts} routine. The background was extracted from a source-free region in the corner of the field of view. Total exposure time of the analyzed observation is 58~ks.

\begin{figure}
\includegraphics[width=0.98\columnwidth]{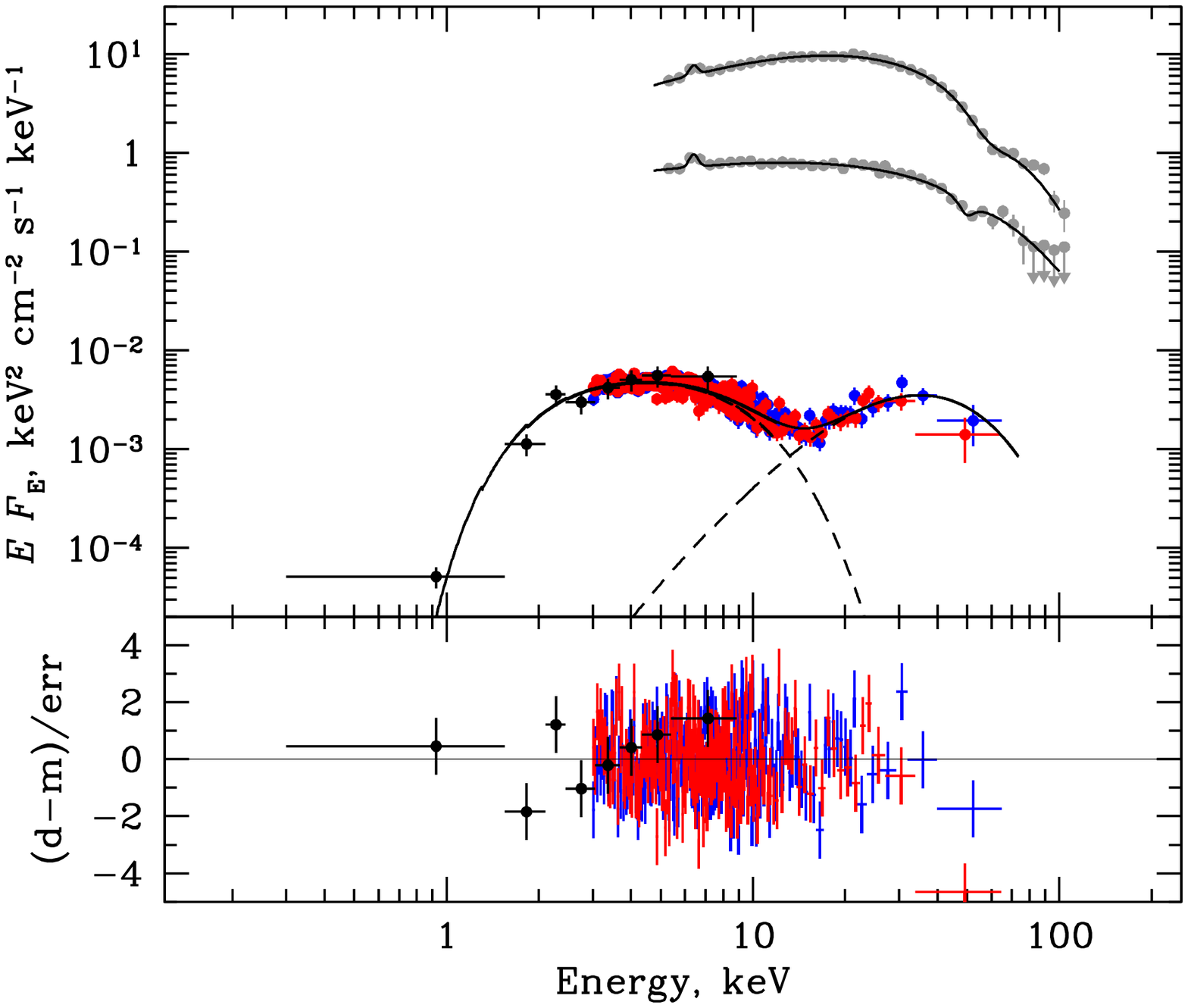}
\caption{$EF_{\rm E}$ spectra of \gx\ in different states with luminosity varying by more than three orders of magnitude, from $2.2\times10^{37}$ \lum\ and $0.2\times10^{37}$ \lum\ for the two  top spectra down to $\sim10^{34}$ \lum\ for the bottom one. 
The bright state spectra (grey circles) are taken from \citet{2015MNRAS.454.2714M}. 
The black, red and blue crosses correspond to the {\it Swift}/XRT and {\it NuSTAR} FPMA and FPMB data in the low state of \gx, respectively. 
The solid line represents the best-fitting model consisting of two {\sc comptt} components shown with dashed lines. 
The corresponding residuals are presented at the lower panel. 
}\label{fig:specs}
\end{figure}

To expand our spectral analysis to lower energies, we used the data from the XRT telescope \citep{2005SSRv..120..165B} on-board the {\it Neil Gehrels Swift Observatory} \citep{2004ApJ...611.1005G} obtained simultaneously with {\it NuSTAR} (ObsID 00088780001). 
Observation was performed in Photon Counting (PC) mode with total exposure of 2~ks.  Spectrum extraction was done using the online tools
\citep[][]{2009MNRAS.397.1177E}\footnote{\url{http://www.swift.ac.uk/user_objects/}}
provided by the UK Swift Science Data Centre. 
To fit the spectra in the {\sc xspec} package we binned the spectra to have at least 1 count per energy bin and fitted them using W-statistic
\citep{1979ApJ...230..274W}.\footnote{\url{https://heasarc.gsfc.nasa.gov/xanadu/xspec/manual/XSappendixStatistics.html}} 
The data from {\it Swift}/XRT and {\it NuSTAR} were used in the 0.3--10 keV and 3--79~keV bands, respectively.

The broadband spectrum of the source as observed by \textit{NuSTAR} and \textit{Swift}/XRT at luminosity around $10^{34}$ \lum\ is shown in Fig.~\ref{fig:specs} (lower data points). 
The spectrum shows two humps around 5 and 40 keV of approximately same fluxes. 
This is very different from the  {\it INTEGRAL} spectra at higher luminosities of $2.2\times10^{37}$ and $0.2\times10^{37}$ \lum (upper two sets of data points in Fig.~\ref{fig:specs}, \citealt{2015MNRAS.454.2714M}), which can be well  fitted with the standard XRP power-law model modified by a high energy cutoff and a cyclotron absorption line at $\sim60$ and $\sim50$ keV, respectively.

\begin{table}
        \begin{center}
        \caption{Best-fitting results for the \gx\ broadband spectrum obtained with {\it NuSTAR}+{\it Swift}/XRT. }
        \label{tab:spe}
        \begin{tabular}{lcc}
\hline
\hline
  Parameter$^{a}$  & Low-energy part & High-energy part \\
\hline
\multicolumn{3}{c}{{\sc comptt+comptt}}\\
\hline
$T_0$, keV           &  \multicolumn{2}{c}{$0.46\pm0.07$} \\[1ex]
$T_{\rm p}$, keV     &  $1.72_{-0.08}^{+0.05}$ &  $8.7_{-0.4}^{+0.9}$ \\[1ex]
$\tau_{\rm p}$        &  $8.4_{-0.9}^{+1.3}$ &  $\gtrsim10$ \\[1ex]
$F_{\rm X}$, $10^{-12}$ \flux  &  $13.3_{-0.6}^{+0.3}$ &  $7.2_{-0.5}^{+1.0}$ \\[1ex]
C-statistic (d.o.f.)   &  \multicolumn{2}{c}{1813.3 (1875)} \\
\hline
\multicolumn{3}{c}{{\sc comptt+gau}}\\
\hline
$T_0$, keV          &  $0.45\pm0.07$ & \\[1ex]
$T_{\rm p}$, keV    &  $1.72_{-0.07}^{+0.08}$ &  \\[1ex]
$\tau_{\rm p}$       &  $8.3\pm0.5$ & \\[1ex]
$F_{\rm X}$, $10^{-12}$ \flux   &  $13.2_{-0.3}^{+0.2}$ & $6.6_{-0.5}^{+0.7}$ \\[1ex]
$E_{\rm gau}$, keV      &   &  $18_{-5}^{+3}$ \\[1ex]
$\sigma_{\rm gau}$, keV &   &  $16_{-2}^{+3}$ \\[1ex]
C-statistic (d.o.f.)   &  \multicolumn{2}{c}{1810.3 (1875)} \\
\hline
\multicolumn{3}{c}{{\sc comptt+gabs}}\\
\hline
$T_0$, keV          &  \multicolumn{2}{c}{$0.63\pm0.04$}  \\[1ex]
$T_{\rm p}$, keV    &  \multicolumn{2}{c}{$15\pm6$}   \\[1ex]
$\tau_{\rm p}$       &  \multicolumn{2}{c}{$1.6_{-1.5}^{+0.9}$}  \\[1ex]
$E_{\rm gabs}$, keV      &     \multicolumn{2}{c}{$14.9\pm0.3$} \\[1ex]
$\sigma_{\rm gabs}$, keV &     \multicolumn{2}{c}{$4.9_{-0.5}^{+0.6}$} \\[1ex]
Depth$_{\rm gabs}$   &    \multicolumn{2}{c}{$13.3_{-2.4}^{+3.5}$} \\[1ex]
$F_{\rm X}$, $10^{-12}$ \flux  &    \multicolumn{2}{c}{$25.7_{-1.1}^{+1.8}$} \\[1ex]
C-statistic (d.o.f.)   &  \multicolumn{2}{c}{1826.4 (1877)} \\
\hline
        \end{tabular}
	\begin{tablenotes}
        \item $^{a}$ Here $T_{\rm p}$, $\tau_{\rm p}$ and $T_0$ are the plasma temperature, plasma optical depth, temperature of the seed photons for the {\sc comptt} model. Fluxes are given  in the 0.5--100 keV energy range.  
        The hydrogen column density is fixed at the Galactic value in the source direction $N_{\rm H}=1.1\times10^{22}$~cm$^{-2}$.
        \end{tablenotes}
\end{center}
\end{table}

To fit the low-state spectrum we explored several models representing different combinations of physically motivated and phenomenological components. 
Particularly, to model the second hump in the spectrum as a separate emission component, we used (i) a combination of two
Comptonization components \citep[{\sc comptt+comptt} model in {\sc
    xspec,}][]{1994ApJ...434..570T}; (ii) a combination of Comptonization for the low-energy hump and an emission line with Gaussian profile for the high-energy one ({\sc comptt+Gau} model in {\sc xspec}).
    Alternatively, the spectrum can be described with a single-component continuum
modified by an absorption feature around 15~keV. In particular, 
we used {\sc comptt+gabs} combination.
Another possibility is to consider bulk Comptonization model {\sc comptb} \citep{2008ApJ...680..602F} to describe the whole continuum, however, we note that the effect of bulk motion should be negligible because of the expected small optical depth of the infalling gas at small accretion rates (see Sect.~\ref{sec:nature_high}).  
We emphasize that parameters of the used Comptonization models have physical meaning only for low magnetic fields and they should be treated with caution for the high-field  XRPs. 

We also included {\sc phabs} component to account for interstellar photoelectric absorption with $N_{\rm H}$  fixed at  the Galactic value in the source direction of $1.1\times10^{22}$~cm$^{-2}$ \citep{2005A&A...440..775K}, because neither strong absorption was ever reported for this source, nor was required by our fit.
To account for minor differences in the absolute flux calibration of the FPMA and FMPB instruments on-board
\textit{NuSTAR} and {\it Swift}/XRT telescope, the normalization factors were added to the separate data groups. 
As a result the fluxes from FPMA and FMPB instruments match to within 2--3 per cent, whereas the XRT normalization is about 1.3 times lower likely due to the fact that \textit{NuSTAR} and {\it Swift} observations were not strictly simultaneous.  
In Fig.~\ref{fig:specs} the fit with two Comptonization components is shown with the corresponding residuals.

The fit results are presented in Table~\ref{tab:spe} and can be summarized as follows: (i) the spectrum can
be fitted equally well with all sets of models; (ii) in the case of two separate spectral components, the flux in the high-energy part is about half of the low-energy one; (iii) in the case of two Comptonization components, the plasma temperature of the high-energy component is about 5 times the low-energy one.

\section{Discussion}
\label{sec:discus}

\begin{figure}
\centering
\includegraphics[width=0.98\columnwidth]{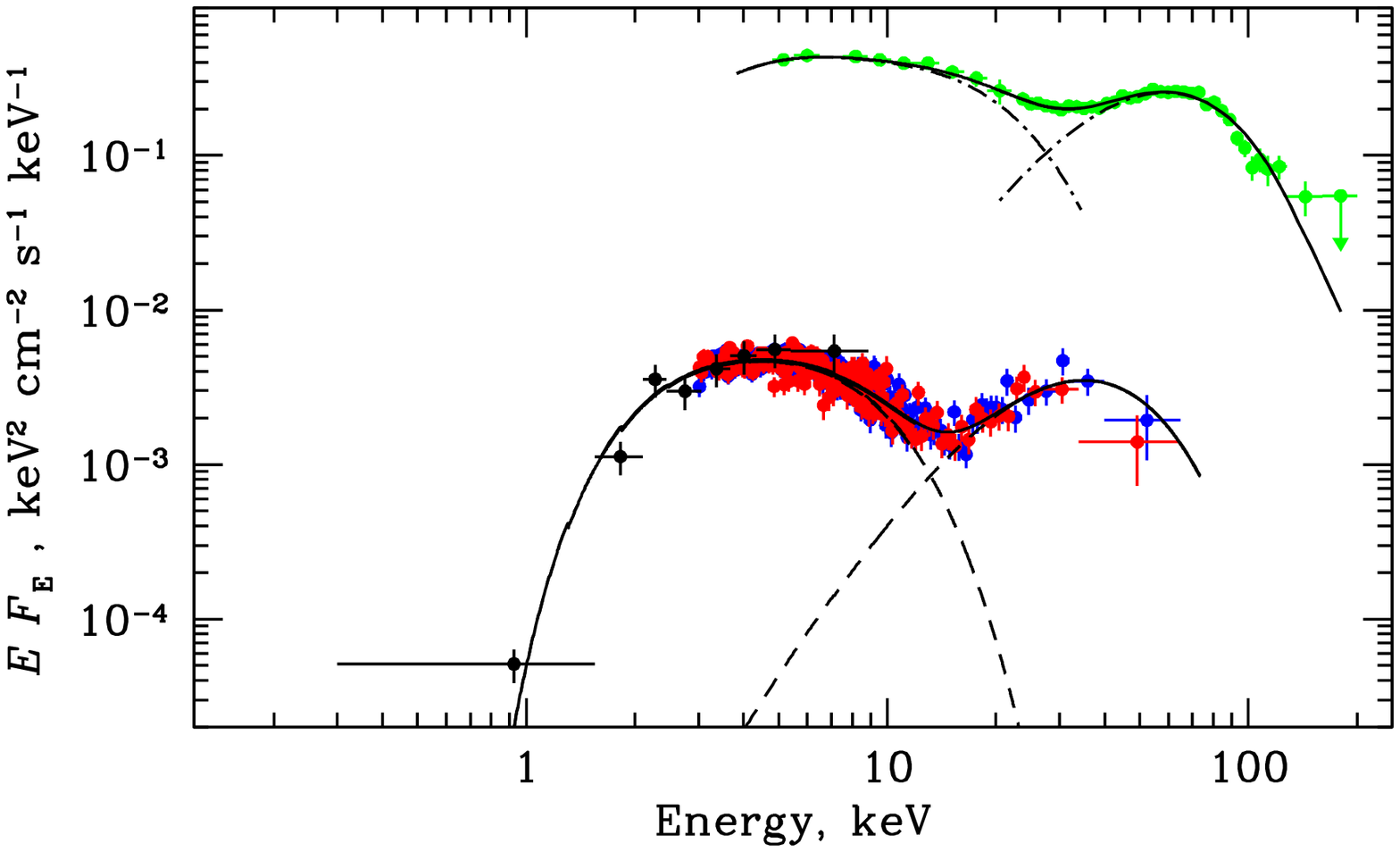}
\caption{The spectra of X~Persei as seen by the {\it INTEGRAL} observatory (green points; \citealt{2012A&A...540L...1D}) and \gx\  
along with the best-fitting models consisting of two Comptonization components.
 Separate model components ({\sc comptt} in {\sc xspec}) are shown with dashed and dash-dotted lines for \gx\ and X~Persei, respectively.}\label{fig:persspec}
\end{figure}

The spectrum of \gx\ in the low-luminosity state resembles the spectrum of another Be/XRP X~Persei obtained at only slightly higher luminosity, $L_{\rm X}\sim8\times10^{34}$~\lum\ (see Fig.~\ref{fig:persspec}). Both sources exhibit double-hump spectra with components peaking at energies separated by a factor of about 5--10. This similarity suggests that physical processes responsible for the spectrum formation at low mass accretion rates are also similar. 
In this section we discuss the possible nature of the emission from \gx\ in the low state and the reasons for the observed abrupt changes of the source spectrum.

\subsection{Cyclotron line}

As already mentioned, a cyclotron line is observed in the spectrum of  \gx\ at high luminosities \citep{2015A&ARv..23....2W}. 
Moreover, a positive  correlation of the line energy with the luminosity has been reported \citep{2012A&A...542L..28K}. 
Particularly, the energy was found to decrease from $\sim60$\,keV at luminosity of $L=2.2\times10^{37}$ \lum\ to $\sim50$~keV at ten times lower luminosity \citep{2015MNRAS.454.2714M}. 
Therefore, one may argue that observed two component spectrum of the source is mimicked by an absorption
feature at $\sim15$ keV between two humps in the source spectrum, which is the cyclotron line shifted from the original $\sim60$ keV due to substantially lower mass accretion rate.

The extrapolation of the observed correlation to low fluxes implies, however, significantly higher energy $E_{\rm cyc} \approx 30$~keV. 
Moreover, such a large decrease appears to be unrealistic in the context of the two models proposed so far to explain the observed correlation of line energy with flux.
In the  model by \citet{2015MNRAS.454.2714M}, the observed correlation  is produced by the Doppler effect in the flow, whose velocity is affected by  radiation pressure, which is more efficient at higher luminosities.  
The maximal relative shift here is limited by a factor of two, as the flow velocity changes from the free-fall with $v\approx 0.5c$ to zero. 
Because the highest observed line energy is $\sim60$~keV, the lowest possible energy is $\sim30$~keV, which is significantly higher than the observed $\sim 15$~keV. 

The second model associates shifts of the CRSF energy with changes of the height of a collisionless shock above the NS polar cap \citep{2017MNRAS.466.2752R}. 
For the dipole field, the decrease of line energy from $\sim$60 to 15~keV implies an increase of  the height of the line forming region by half a NS radius. 
The geometric dilution of radiation from a polar cap of size $<100$~m is very significant at such height, with only $\sim$1 per cent of the cap continuum photons passing through the cross-section of the accretion channel  (which is only 3--4 times larger at this height than the cap area). We conclude, therefore, that the description of the observed low luminosity spectrum of \gx\ with a single continuum component with a CRSF at $\sim15$\,keV is implausible also in this model.

\subsection{Nature of the high-energy component}
\label{sec:nature_high}

The conclusion above implies that the two humps in X-ray spectrum of the source are likely indeed two separate components.
The soft blackbody-like component with a typical temperature of $kT \le 2$\,keV can be explained as radiation from hotspots at the NS surface heated up by the accretion process. 
However, the origin of the second component, which  can also be fitted by a blackbody with the temperature $\approx 9$\,keV or a broad Gaussian emission line  (see Table\,\ref{tab:spe}) is much less clear. Below we discuss several possible physical origins of this spectral component:

\begin{enumerate}
\item {\it Bulk Comptonization.} 
As was mentioned in the Introduction, hard spectral component in X~Persei was described by the model of bulk Comptonization in the accretion channel \citep{2012A&A...540L...1D}. 
We note that the effective interaction between electrons and photons is only possible if optical depth of the interaction region is sufficiently large ($\tau \sim 5-10$).
Such optical depth across the accretion flow is only realized at high accretion rates corresponding to the onset of an accretion column \citep{2015MNRAS.447.1847M}. 
In case of \gx, an upper limit on the Thomson optical depth in the interaction region is about $\tau_{\rm T} \approx 0.01 (L_{\rm X}/10^{34}\,\rm erg\,s^{-1}) \ll1$  \citep[see Eq. (18) in][]{2015MNRAS.447.1847M}.
We conclude, therefore, that bulk Comptonization is likely inefficient in  \gx\ and in other pulsars with $L_{\rm X} \le 10^{35}$\,erg\,s$^{-1}$, i.e. including X~Persei.
This conclusion is also supported by our spectral fit using the {\sc comptb} model. Particularly, fixing efficiency of bulk over thermal Comptonization (parameter $\delta$) at zero value allows us to get an excellent quality fit with parameters similar to the {\sc comptt+comptt} model.

\item {\it  Collisionless shock.} The hard component may originate from the collisionless shock above NS surface \citep{1975ApJ...198..671S,1982ApJ...257..733L}, where 
the kinetic energy of the accreting particles is transferred to their thermal energy with characteristic temperatures for ions and electrons being close to their virial temperatures.
In this case, Coulomb interactions might be effective enough to transfer thermal energy of ions to the electrons which are able to cool effectively by cyclotron radiation, resulting in cooling of the post-shock region and settling of the plasma onto the NS surface.
The height of the post-shock region determined by the cooling rate is comparable to the NS radius and its spectrum  for $\dot M = 6 \times 10^{14}$\,g\,s$^{-1}$ does qualitatively resemble the spectra observed in X~Persei and \gx\ in their low-luminosity state \citep{1982ApJ...257..733L}.  
However, the spectra in this model  were computed under assumption of optically thin cyclotron emission, which is not correct, and the physical mechanism for the formation of a hypothesized collisionless shock is not really clear. 
It is therefore premature to draw any firm conclusions.

\item {\it Cyclotron emission.}
Thermally broadened cyclotron emission as a source of the hard component in X~Persei was discussed by \cite{1998ApJ...509..897D} based on the model by \cite{1993ApJ...418..874N}. 
Because the line forming region in this scenario is assumed to be located at the NS surface, the temperature determining the line broadening was expected to be similar to the temperature of the soft component of about 2 keV, which is much smaller than the observed width of the hard component. 
On the other hand, the upper layers of the NS atmosphere heated by the accretion are not cooled efficiently by their thermal emission and can be overheated to $\sim 10-100$\,keV depending on the accretion rate and the magnetic field strength \citep{2000ApJ...537..387Z,2001A&A...377..955D,2018A&A...619A.114S, 2018MNRAS.tmp.3002G}. 
In this case, thermal broadening of the cyclotron emission line, which forms in these hot layers, is high enough to explain the observed width of the hard component.
However, there is still a strong discrepancy between the model and observations in \gx: the hard component  reaches its maximum at an energy significantly below the cyclotron energy at the NS surface.
We argue that this discrepancy can be explained by specifics of radiative transfer in the atmosphere of highly magnetized NS which is determined by multiple Compton scattering of X-ray photons. 
The scattering cross-section is resonant at the cyclotron energy \citep{2016PhRvD..93j5003M} and photons cannot leave the atmosphere close to this energy. 
Instead, photons diffuse to lower energies by scatterings and leave the atmosphere in a low-energy wing, where the cross-section is small enough. 
Thus, the hard component can be explained by cyclotron emission in the atmosphere heated by accretion and reprocessed by magnetic Compton scattering.

\item {\it Thermal Comptonization.} Existence of the overheated upper layers of the NS atmosphere can explain the hard energy spectral component by the Comptonization of ordinary mode photons (which experience Compton scattering with a larger cross-section in comparison with extraordinary photons).
The examples of Comptonized spectra calculated by \citet{1988Ap.....28..253L} show two components, which should have significant difference in polarization. 
This scenario can be verified with upcoming X-ray polarimeters.
We note that Comptonization of extraordinary mode photons is also expected to contribute to the broadening of the cyclotron line discussed in previous paragraph, i.e.  two mechanisms are highly complementary and likely operate simultaneously. 
These effects should be taken into account in detailed calculations of emerging spectra and polarisation.

\end{enumerate}

\section{Conclusion}

We presented here the analysis of the spectrum of XRP \gx\ observed with \textit{NuSTAR} in a very low state with $L_{\rm X} \sim 10^{34}$\,erg\,s$^{-1}$.
The spectrum was found to differ dramatically from the previously observed spectra of the source at higher luminosities ($L_{\rm X} \sim 10^{36} - 10^{37}$\,erg\,s$^{-1}$). 
The latter are typical for XRPs and can be described by Comptonization in a hot electron slab with a cyclotron absorption feature at 45--60\,keV. 
On the other hand, at low luminosity two components with characteristic temperatures of $kT\sim 2$\,keV and $\sim 10$\,keV
can be identified. A similar spectrum has been reported previously only for the persistent low-luminosity XRP X~Persei.
A transition between the two spectral states has, however, never been observed in any XRP.

Given the similarity in low-luminosity spectra of \gx\ and X~Persei, it is reasonable to assume that the physical processes shaping their spectra are also similar. 
The bulk Comptonization interpretation invoked previously for X~Persei to explain the hard part of the spectrum is, however, problematic for \gx\ due to necessarily low optical depth of the  accretion flow. We have also demonstrated that the two-component spectrum cannot be mimicked by the flux depression between the two components by the resonant scattering.

We  considered two possibilities to explain the observed low-luminosity spectrum of \gx. 
The first scenario is associated with electron heating by hot ions. 
In this case, the softer component is associated with thermal emission from the deep layer of temperature $\sim 2$\,keV, while the harder component is associated with cyclotron photons reprocessed further by magnetic Compton scattering in the upper layers of the atmosphere. 
In the second scenario, the thermal polar cap emission is partly Comptonized in the overheated upper layers of the atmosphere.
Both these models are rather qualitative at this stage and have to be verified by numerical modelling and comparison with the available observations of both sources.

\section*{Acknowledgements}
This work was supported by the grant 14.W03.31.0021 of the Ministry of Science and Higher Education of the Russian Federation.
VD thanks the Deutsches Zentrum for Luft- und Raumfahrt (DLR) and Deutsche Forschungsgemeinschaft (DFG) for financial support.  
VFS acknowledges the support from the DFG grant WE 1312/51-1 and from the travel grant of the German Academic Exchange Service (DAAD, project 57405000). This research was also supported by the 
 Academy of Finland travel grant 317552 (ST, JP) and by the  Netherlands Organization for Scientific Research (AAM). ARE and RW acknowledge the {\it NuSTAR} team for approving the DDT observation of GX\,304$-$1.



\bsp    
\label{lastpage}
\end{document}